\documentclass[%
reprint,
twocolumn,
superscriptaddress,
showpacs,preprintnumbers,
nofootinbib,
aps,
prl,
]{revtex4-1}

\usepackage{subfigure}
\usepackage[colorlinks, linkcolor=blue, anchorcolor=blue, citecolor=blue]{hyperref}
\usepackage{amsmath}
\usepackage{amssymb}
\usepackage{mathtools}
\usepackage{enumerate}
\usepackage{hyperref}
\usepackage{graphicx}
\usepackage{color}

\usepackage{ulem}

\usepackage{comment}

\graphicspath{
  {./}
  {./figures/}
}

\newcommand{\beq}{\begin{equation}}
\newcommand{\eeq}{\end{equation}}
\newcommand{\bea}{\begin{eqnarray}}
\newcommand{\eea}{\end{eqnarray}}
\newcommand{\ba}{\begin{array}}
\newcommand{\ea}{\end{array}}
\newcommand{\bit}{\begin{itemize}}
\newcommand{\eit}{\end{itemize}}

\newcommand{\eq}[1]{Eq.~(\ref{#1})}

\definecolor{purple}{rgb}{0.5,0,0.5}
\definecolor{brown}{rgb}{0.7,0.5,0}
\definecolor{teal}{rgb}{0.2,0.6,0.2}

\begin{document}
\newcommand{\TQ}{\affiliation{
MOE Key Laboratory of TianQin Mission, TianQin Research Center for Gravitational Physics \& School of Physics and Astronomy, 
Sun Yat-sen University (Zhuhai Campus), Zhuhai 519082, China.
}}

\title{Cosmological first-order phase transitions without bubbles}

\author{Dongdong~Wei}
\email{weidd5@mail2.sysu.edu.cn}
\author{Haibin~Chen}
\author{Qiqi Fan}
\author{Yun~Jiang}
\email{jiangyun5@sysu.edu.cn (corresponding author)}
\TQ

\date{\today}

\begin{abstract}
In the traditional view a cosmic first-order phase transition cannot occur without 
nucleating handful of bubbles in the entire Hubble volume.
The presence of domain walls during the transition may, however, significantly alter the dynamics of the phase transitions. Using lattice simulation, we demonstrate that vacuum fluctuations induce the destabilization of the domain walls that will classically transform into the domain trenches of the true vacuum, resulting in successful phase transitions without bubbles. After providing an analytical method to estimate the temperature at which the domain trenches are produced, we take the $\mathbb{Z}_2$-odd singlet model as an example and conclude that the bubble-free mechanism developed in this Letter constitutes a competing means of completing the phase transition against with quantum tunneling, opening up the new viable parameter region. 
\end{abstract}

\maketitle

\emph{Introduction.} 
First-order phase transition (FOPT) is of particular interest for cosmology as it provides a solution to the observed baryon asymmetry of the early Universe~\cite{Morrissey:2012db} and leads to the generation of gravitational waves (GWs) that is potentially detectable at GW detectors~\cite{Caprini:2019egz, TianQin:2020hid, Liang:2021bde, Cai:2017cbj}.
In some models, it is also related to the problem of dark matter production, see, e.g.,~\cite{Baker:2016xzo, Baker:2019ndr}. 

In a FOPT, there exists a critical temperature $T_c$ at which two degenerate free-energy minima are separated by a potential barrier, and below which this energy degeneracy is broken due to thermal effects. 
The lower minimum corresponds to the true vacuum of the theory, while the higher one is called the false vacuum, which is quantum-mechanically metastable. 
If the Universe experiences a FOPT, the (high-temperature) phase at $T>T_c$ will play the role of the false vacuum, 
and the transition to the true vacuum (also say the false vacuum decay) proceeds via nucleating bubbles of the true vacuum, as a result of tunneling process~\cite{Coleman:1977py} or due to thermal fluctuations large enough to jump over the barrier~\cite{Borrill:1994nk}. 
In the cosmological context, the nucleation of bubbles is efficient when the probability of the bubble nucleation per Hubble volume per Hubble time is roughly of order one~\cite{Anderson:1991zb}, otherwise the phase transition will not happen and the universe will be trapped in the false vacuum. This will cause a catastrophic inflation~\cite{Guth:1982pn, Hawking:1982ga} and if the true vacuum is the desired phase at $T<T_c$, (i.e., the electroweak (EW)-broken minimum accounting for a proper EW symmetry breaking~\cite{Baum:2020vfl,Biekotter:2021ysx,Goncalves:2021egx,Biekotter:2022kgf}, then perhaps a {\it bubble-free} phase transition is crucially needed.

In addition, phase transitions are responsible for the possible formation of topological defects in the early Universe~\cite{Vilenkin:2000jqa}.
In field theory, domain wall, as one of the defects, arises from spontaneous breaking of a discrete symmetry\footnote{Such discrete symmetries occur frequently in many particle physics models.}, 
resulting in degenerate vacua. 
In the case that such a symmetry breaking occurred in the early universe, the universe may end up with different vacuum domains divided by domain walls~\cite{Vilenkin:2000jqa}. 
The existence of domain walls will give rise to a rich impact on the dynamics of the cosmological phase transition.
For instance, domain walls can act as impurities to catalyze bubble nucleation, thereby enhancing the tunneling probability at the location of the domain walls to complete the FOPT~\cite{Blasi:2022woz, Blasi:2023rqi, Agrawal:2023cgp}. 
However, this realization is still based on the quantum tunneling effect and largely relies on the considerable area occupied by the domain walls, which is not yet verified in the entire parameter space. Instead, in this {\it Letter} we suggest a competing way to achieve the FOPTs where bubble nucleation is inefficient or even absolutely prohibited. 
We find that the domain walls can be spontaneously destroyed by virtue of vacuum fluctuations and classically transform into the true vacuum state. We term it the {\it domain trench}, which will, in turn, play the role of vacuum bubbles and absorb the initial false vacuum domains in the subsequent process, leading to a successful phase transition to the true vacuum. 

It is worth pointing out that in our mechanism the “domain wall problem”~\cite{Zeldovich:1974uw} encountered by most models with discrete symmetries can be safely escaped and the trapped false vacuum can also be successfully rescued, both of which have important cosmological significance.

\emph{Model}. 
Generally speaking, a phase transition consisting of two steps is the minimal realization of this new approach. The first step is responsible for the formation of domain walls, will make the FOPT occur in the second step. 
As a simple model admitting the domain wall solution and a two-step phase transition, we consider the SM extended with a real scalar singlet $S$, which is odd under a $\mathbb{Z}_2$ symmetry. 
The effective potential at finite temperature in terms of the two scalar condensates, $h= \sqrt{2}\langle H \rangle_T$,  $s=\langle S \rangle_T$\footnote{At $T=0$, $s=0$ and $h=v_{\rm EW}\simeq 246~{\rm GeV}$ required by the proper electroweak symmetry breaking.} takes the form,

\bea
\label{eq:potent}
V(h,s,T) &=&-\frac{1}{2} \left( \mu_h^2- c_h T^2\right) h^2+\frac{1}{4} \lambda_h h^4 \label{potential}\\
&&-\frac{1}{2}\left( \mu_s^2-c_s T^2\right) s^2+\frac{1}{4} \lambda_s s^4+\frac{1}{2} \lambda_{h s} h^2 s^2 \nonumber,
\eea
where $\mu^2_h$ and $\lambda_{h}$ are the parameters fixed by the electroweak scale $v_{\rm EW}$ and the observed Higgs mass $m_h=125 {\rm~GeV}$~\cite{CMS:2022dwd, ATLAS:2022vkf}, and $c_h$, $c_s$ are the terms arising from the thermal masses~\cite{Espinosa:2011ax}.

\begin{figure}[t] 
\centering 
\includegraphics[width=0.45\textwidth]{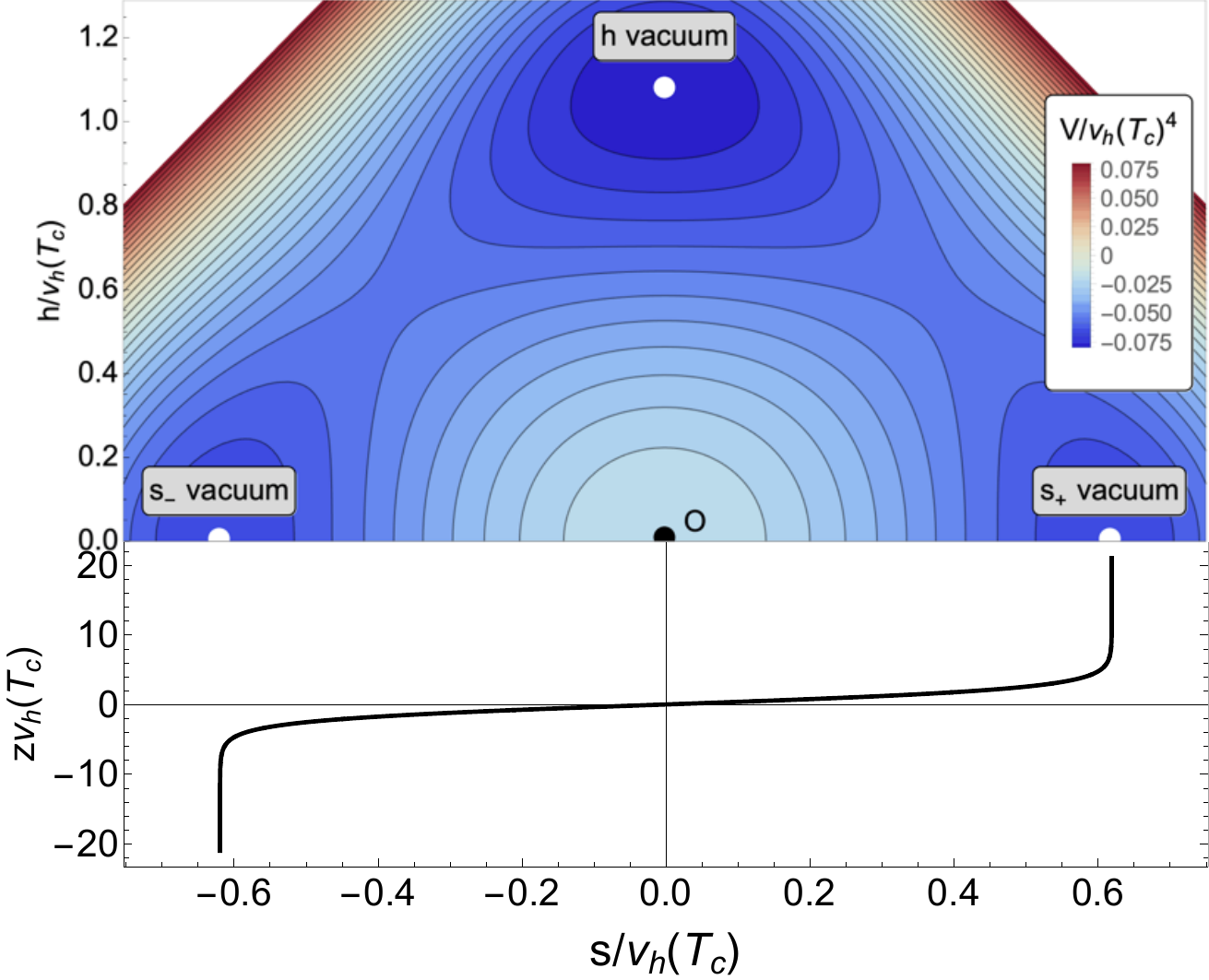} 
\caption{The potential and the associated static domain wall solution at $T=75 {\rm~GeV}<T_c$ for the BMP-A. There exists a barrier between $s_\pm$ vacuum and $h$ vacuum as a feature of the FOPT.} 
\label{fig:potent}
\end{figure}

For the formation of domain walls, we require the spontaneous breaking of the $\mathbb{Z}_2$ symmetry in the first step.
At $T=T_{\rm dw}=\sqrt{\mu_s^2/c_s}$, the minimum of the potential starts departing from the origin and develops in the $s$ field direction at $s=\pm v_s(T), v_s(T)=\sqrt{(\mu_s^2-c_s T^2)/\lambda_s}$, leading to the coexistence of two energetically equivalent domains: $s_\pm$ domain with $(h,s)=(0,\pm v_s)$. 
In the transition region between $s_\pm$ domains, domain walls are generated as a result of the smooth interpolation of the field configurations. For a static planar domain wall in the x-y plane at $z = z_0$, the field configurations interpolating between the $s_\pm$ vacuum are given by~\cite{Vilenkin:2000jqa}
\beq
\label{eq:solution_dw}
\begin{aligned}
s_{\rm dw}(\mathbf{x},z_0,T)=&~v_s(T) \tanh [(z-z_0)/L_{\rm dw} (T)],\\
h_{\rm dw}(\mathbf{x},T)=&~0,
\end{aligned}
\eeq
where the thickness of the wall is characterized by $L_{\rm dw} (T)=  \sqrt{2/(\mu_s^2- c_s T^2)}$. 

For the scenario of our interest, at $T = T_c<T_{\rm dw}$ a new minimum having the same value of $V$ as the $s_\pm$ vacuum appears in the $h$ field direction at $h=v_h(T_c), v_h(T)=\sqrt{(\mu_h^2-c_h T^2)/\lambda_h}$, indicating a FOPT in the second step. When the temperature drops below $T_c$ (as illustrated in Fig.~\ref{fig:potent}), this $h$ vacuum develops into the true vacuum and meanwhile, the $s_\pm$ vacuum becomes the false vacuum,  between them there exists a barrier. 

Two benchmark points (BMPs) predicting a two-step phase transition described above are provided in Table~\ref{tab:BMP}. 
Due to the failure to homogeneous bubble nucleation, neither point can lead to a successful phase transition.
The presence of domain walls during the transition may, however, alter the result, which we shall demonstrate.\footnote{For the BMP-A inhomogeneous bubbles seeded by domain walls can also play the role of catalyzing the phase transition~\cite{Blasi:2023rqi}. The competition with the new mechanism developed in this work will be clarified later.} This is essentially because there is a higher-energy state (compared to the $h$ vacuum) in the domain wall that coexists with the $s_\pm$ vacuum. 

\begin{table}[t]
\centering
\caption{Two BMPs admitting the two-step phase transition with inefficient nucleation of homogenous bubbles. 
$T_c=\sqrt{\frac{-c_h \lambda_s \mu_h^2+c_s \lambda_h \mu_s^2 +\sqrt{\lambda_h \lambda_s(c_s \mu_h^2-c_h \mu_s^2)^2}}{\lambda_h c_s^2-\lambda_s c_h^2}}$ and the mass squared of the physical singlet $m_s^2=\lambda_{hs}v_{\rm EW}^2-\mu_s^2$. 
All the dimensional quantities are in the unit of GeV or ${\rm GeV}^{-1}$.}
\vspace{1mm}
\begin{tabular}{ c | c c c c c c c c } 
 \hline
 \hline
 BMP & $\lambda_s$ & $\lambda_{hs}$ & $m_s$  & $T_{\rm dw}$  & $T_c$  &$v_s(T_{c})$&$v_h(T_{c})$ & $L_{\rm dw}(T_c)$\\ [0.5ex] 
 \hline 
A &1 & 0.73    & 168     &  186 &85  &113 & 186 &0.01\\  
B &1 & 0.86    & 181     & 203   & 56 &133 &222 &0.01\\
 \hline
 \hline
\end{tabular}
\label{tab:BMP}
\end{table}

\emph{Numerical simulation of the wall instability}.
After formation, the domain walls start the evolution, which is governed by the equations of motions of $\phi=h, s$ fields,
\beq
\label{eq:eom}
\ddot{\phi}-\nabla^2 \phi+\frac{\partial V(h,s)}{\partial \phi}=0.
\eeq

Limited by the dynamical range, we perform the lattice simulation from $T_c$ in a volume much smaller than the Hubble size, $H^{-1}(T_c)$. 
The key issue is the initialization of the field configuration. 
In fact, the size of the domain walls at $T_c$ depends on the strength of the interactions involving the $s$-field. 
Assuming that the relevant couplings, $\lambda_s$ and $\lambda_{hs}$, of our BMPs are sufficiently weak, 
the domain walls will quickly reach the scaling regime and can be stretched to the curvature radius that is comparable to $H^{-1}(T_c)$~\cite{Saikawa:2017hiv}.\footnote{For the case of strong couplings, it is difficult to stretch out the initial wall segments, resulting in the typical curvature radius of walls at $T_c$ much smaller than $H^{-1}(T_c)$~\cite{Vilenkin:2000jqa}. In this situation using the planar solution to describe the domain walls may not be viable.}
In this case, it is safe to assume that the domain walls at $T_c$ have good planar symmetry in the simulation so that they can be described by the static field solution, \eq{eq:solution_dw}. 
Following~\cite{Braden:2018tky}, we set the initial profile for the two fields as follows:
\bea
&s(\mathbf{x}, t=0)&=s_{\rm dw}, \quad \dot{s}(\mathbf{x}, t=0)=0, \label{eq:initials} \\
&h(\mathbf{x}, t=0)&=\delta h(\mathbf{x},\tau_{\rm ref}), \quad \dot{h}(\mathbf{x}, t=0)=\delta \dot{h}(\mathbf{x},\tau_{\rm ref}),
\label{eq:initialh}
\eea
where $\delta h$ and $\delta{\dot h}$  are uncorrelated and spatially inhomogeneous perturbations that originate from the vacuum fluctuation on the $h$ field.\footnote{We neglect the perturbation of $s$ field as it contributes to the high-order corrections to \eq{eq:eom}.} In the quantum vacuum state, at any instant $\tau$, the Fourier-transformed modes $\delta {\tilde h}_{\bf k}$ do not have well-defined values but each component has a probability distribution~\cite{Liddle:2000cg}.\footnote{In fact, the vacuum state can be expanded in terms of states in which they do have well-defined values, and the probability of finding a given set of values is the modulus-squared of $|\delta {\tilde h}_{\bf k}|^2$ in the expansion. According to quantum field theory, the squared vacuum fluctuation $\sigma^2_k$ corresponds to the expectation value of $\delta h\delta h$ on the vacuum, $\langle \delta h\delta h \rangle = (2\pi)^{-3} \int \!d^3 k  |\delta\tilde{h}_{\bf k}^2|$.} 
Since $\delta {\tilde h}_{\bf k}$, for a given ${\bf k}$, follows the dynamics of a harmonic oscillator (c.f. \eq{eq:lineom}), in the vacuum sate it obeys the stochastic Gaussian distribution~\cite{Liddle:2000cg},
 \beq
 \label{eq:distr}
 \mathcal{P}\left(\delta {\tilde h}_{\bf k}(\tau)  \right) = \frac{1}{\sqrt{2\pi \sigma^2_k}} \exp \left(- \frac{\left| \delta {\tilde h}_{\bf k}(\tau) \right|^2} {2\sigma^2_k}\right),
 \eeq
 where the variance of the probability distribution $\sigma^2_k$, where $k\equiv |{\bf k}|$, is independent of the direction of ${\bf k}$ and is given by the equal-time correlation function
 \beq
\langle \delta {\tilde h}_{\bf k}^*(\tau) \delta {\tilde h}_{\bf p}(\tau)\rangle=\sigma^2_k (2\pi)^3 \delta^{(3)}({\bf k}-{\bf p}),
\eeq
leading to $\sigma^2_k=(2\omega_{\bf k})^{-1}$ in the flat metric.

\begin{figure}
 \centering
 \subfigure
 {
  \begin{minipage}[t]{0.12\textwidth}
   \centering
   \includegraphics[width=2.3cm]{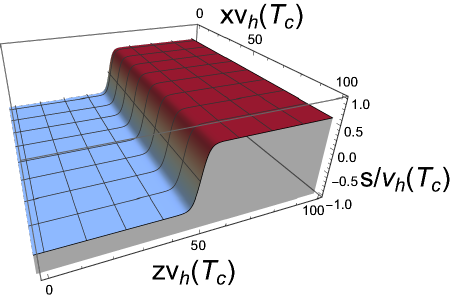}
  \end{minipage}
  \begin{minipage}[t]{0.12\textwidth}
   \centering
   \includegraphics[width=2.3cm]{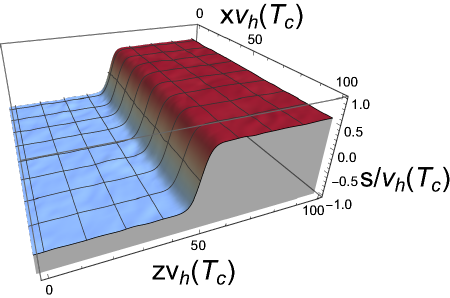}
  \end{minipage}
  \begin{minipage}[t]{0.12\textwidth}
   \centering
   \includegraphics[width=2.3cm]{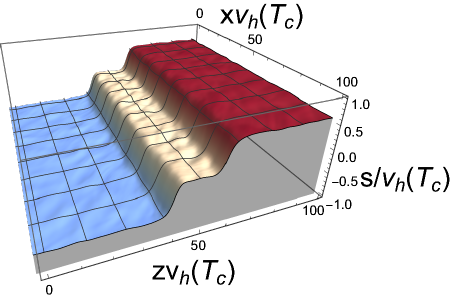}
  \end{minipage}
  \begin{minipage}[t]{0.12\textwidth}
   \centering
   \includegraphics[width=2.3cm]{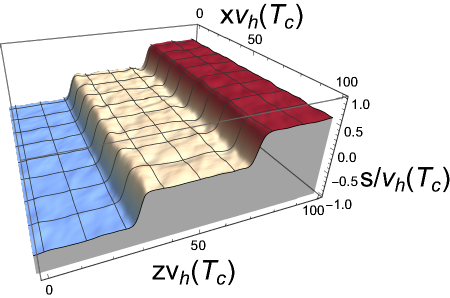}
  \end{minipage}
 }
 \subfigure
  {
  \begin{minipage}[t]{0.12\textwidth}
   \centering
   \vspace{-0.3cm}
   \includegraphics[width=2.3cm]{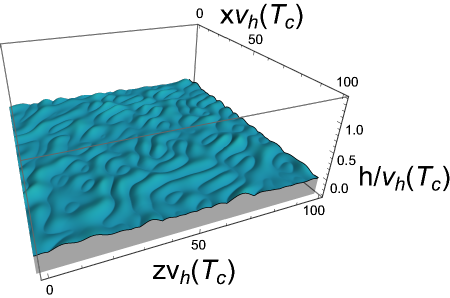}
  \end{minipage}
    \begin{minipage}[t]{0.12\textwidth}
   \centering
      \vspace{-0.3cm}
   \includegraphics[width=2.3cm]{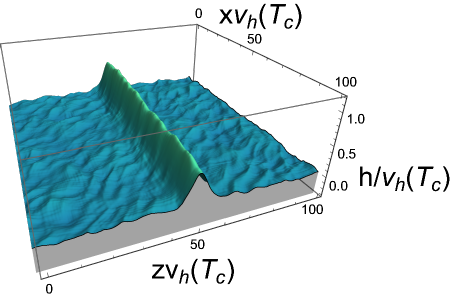}
  \end{minipage}
    \begin{minipage}[t]{0.12\textwidth}
   \centering
      \vspace{-0.3cm}
   \includegraphics[width=2.3cm]{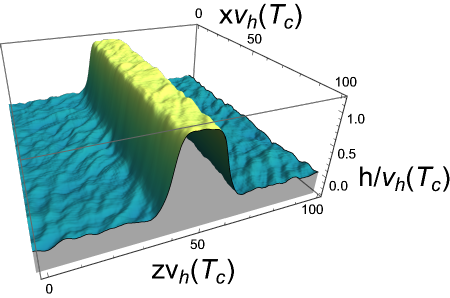}
  \end{minipage}
    \begin{minipage}[t]{0.12\textwidth}
   \centering
      \vspace{-0.3cm}
   \includegraphics[width=2.3cm]{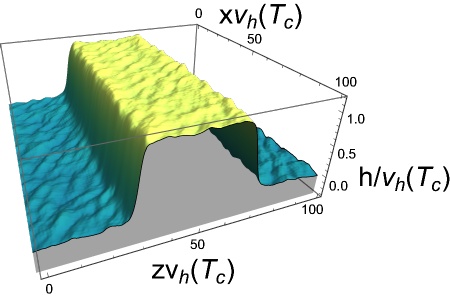}
  \end{minipage}
 }
\caption{Dynamical evolution of a planar wall existing in space between the $s_+$ (red region) and $s_-$ (blue region) domains in the presence of an inhomogeneous fluctuation. The simulation is performed at $T=75{\rm~GeV}<T_c$. 
We present the configurations for the $s$ (upper panel) and $h$ (lower panel) fields in the x-z plane at $tv_{h}(T_c) =0, 20, 40, 54$ (from left to right).} 
\label{fig:IFres} 
\end{figure}

We set up a pair of domain walls at $z_0 =\pm L/4$, so the separation of the nearby domain walls is $R= L/2$. 
Following the description given in Appendix B (see also~\cite{Braden:2018tky}), we initialize the field configurations, \eq{eq:initialh} and perform the lattice simulation where the Crank-Nicholson leapfrog algorithm~\cite{Press:1989yh} is utilized for generating the time evolution of the dynamical fields. 
All the simulations are terminated by $t_{\rm end}/R=1$ during which the interference effect from the nearby domain wall does not arise and the temperature change is also negligible. 
A striking example using the BMP-A  is provided by Fig.~\ref{fig:IFres}. 
Until at $t v_{h}(T_c)=20$ (second column), the $h$ field in the wall region ($z=z_0$)  starts shifting towards the value $h=v_h$ of the $h$ vacuum but the domain wall remains stable until the $s$-field configuration spreads out.
About $t v_{h}(T_c)=40$ (third column) the domain wall destabilizes and quickly turns into the domain trench, which subsequently grows wider, making the entire volume eventually transition to the $h$ vacuum state. 
Therefore, our numerical results clearly demonstrate that the inhomogeneous vacuum fluctuations can cause the instability of the domain wall and the production of the domain trench to complete the FOPT without bubble nucleation.

\begin{figure}[t]
\centering 
\includegraphics[width=0.5\textwidth,height=8cm]{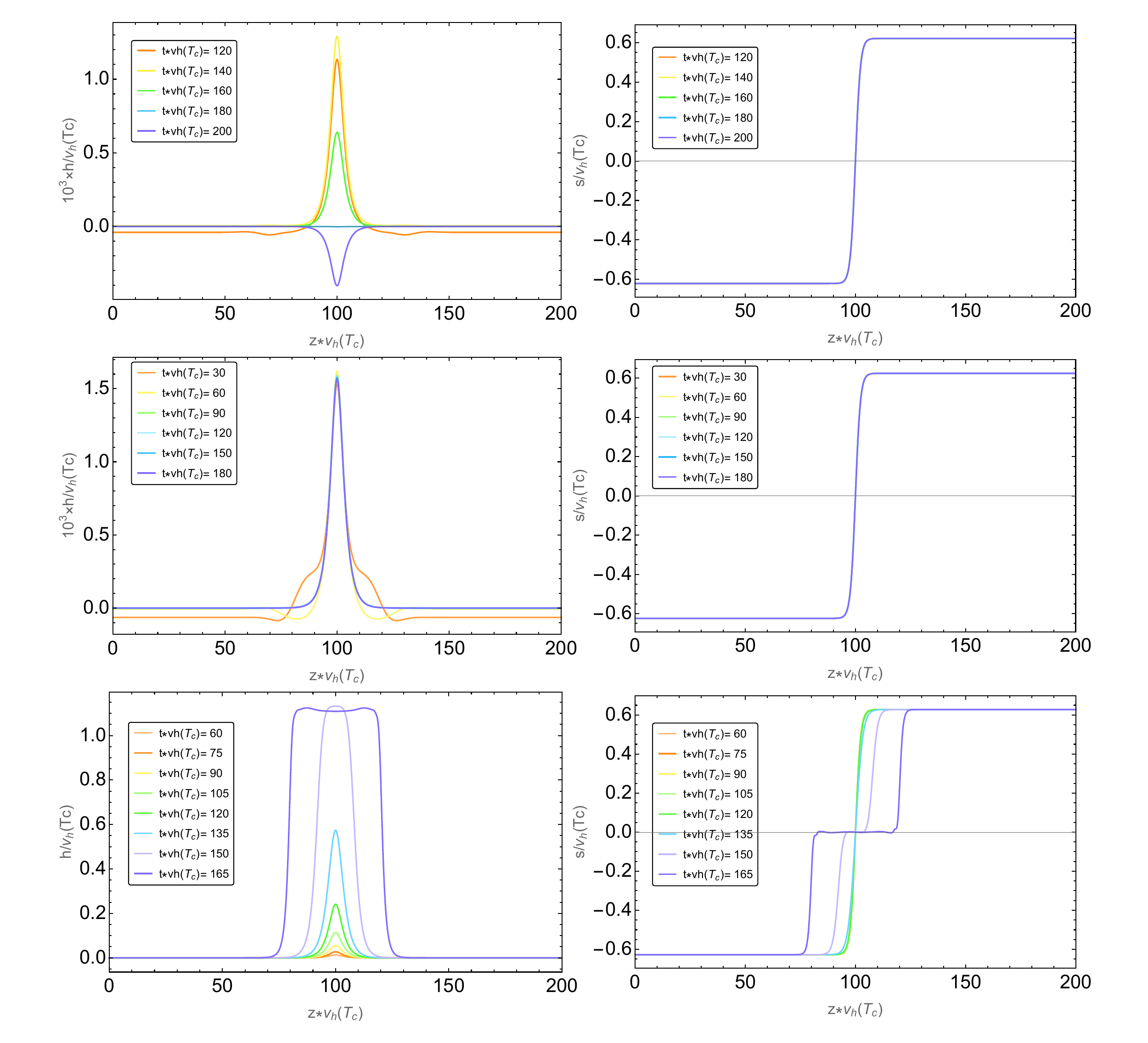} 
\caption{Time evolution of a pair of planar walls centered at $z_0=\pm100/v_h(T_c)$ with a spatially homogeneous fluctuation $\Delta=5\times 10^{-3}$ at $T>T_{\rm res}, T\simeq T_{\rm res}, T<T_{\rm res}$ (from top to bottom). The $h$ and $s$ field configurations are shown in the left and right columns, respectively. The negative $z$ axis is omitted due to symmetric configuration.} 
\label{fig:HFevol}
\end{figure}

\emph{Methods of evaluating the rescue temperature}. 
While the inhomogeneous fluctuations can be implemented in the simulations, we are unable to evaluate from which temperature the domain wall becomes the domain trench. 
To circumvent this trouble, we consider the homogeneous (spatial-independent) fluctuations $\delta h \equiv \Delta$ and neglect the effect of $\delta \dot{h}$ in the simulation. 
In Fig.~\ref{fig:HFevol} we present the time evolution of the field configurations at three distinct temperatures below $T_c$. For instance, 
at $T\simeq T_c$ (upper panel) we see that only the $h$ field exhibits a small oscillation around $h=0$, but the $s$ field has no significant change, so the domain wall is stable at this time.
On the contrary, at $T\ll T_c$ (lower panel) the $h$ field in the wall region will quickly move to $h=v_h$ and meanwhile, the $s$ field will tend to zero. 

\begin{figure}[t]
\centering 
\includegraphics[width=0.4\textwidth]{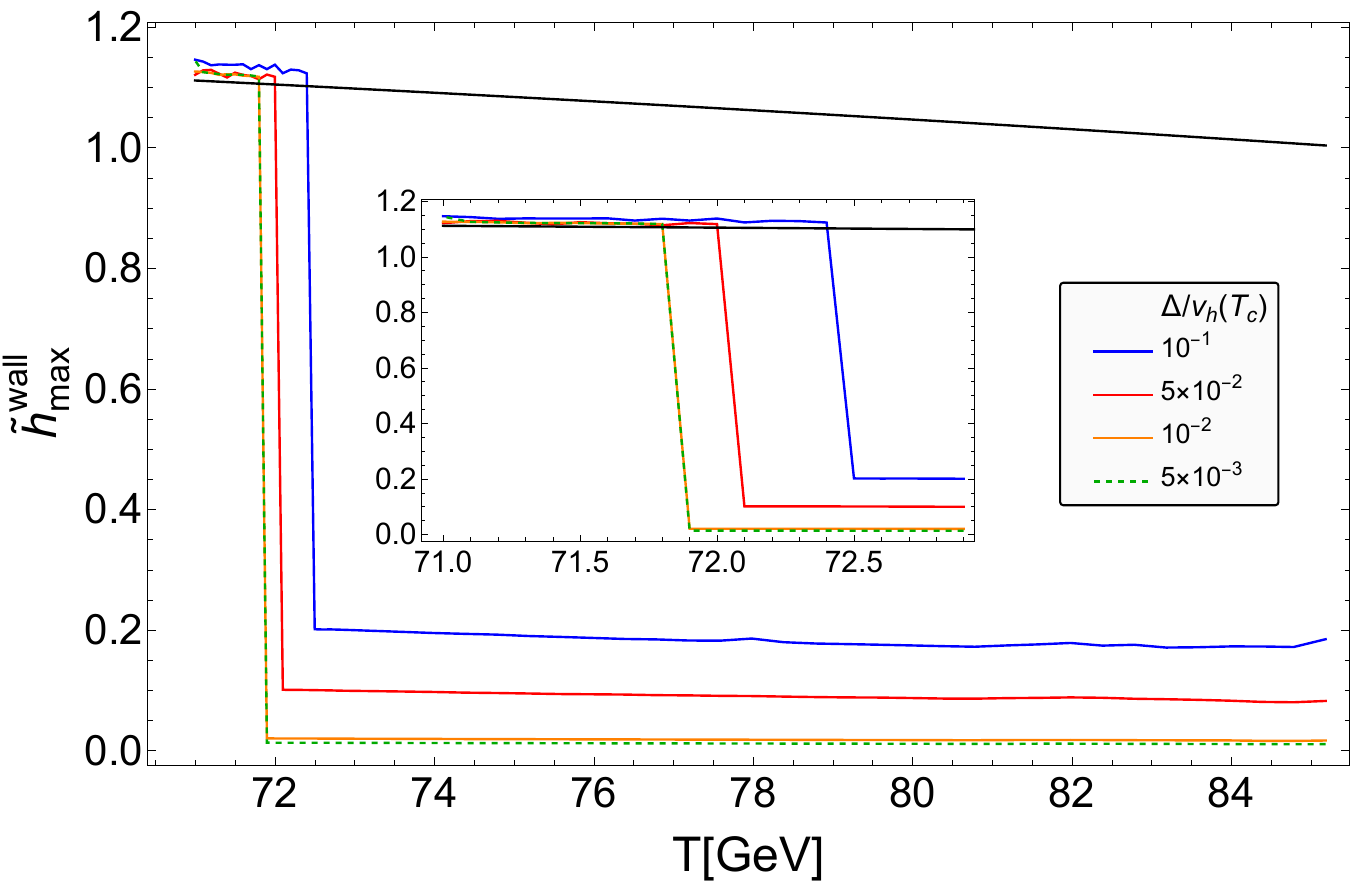} 
\caption{The value of $\tilde{h}^{\rm wall}_{\rm max}(T)$ changes in $T$ starting from $T_c$ in the presence of homogeneous fluctuations. The black line shows the ratio $v_h(T)/v_h(T_c)$, which equals to one at $T=T_c$. From its intersection with the other lines, we can determine $T_{\rm res}$ under different values of $\Delta$.} 
\label{fig:maxvalue} 
\end{figure}

By finely adjusting $T$ in between them, we find a critical temperature at which a stable configuration can be achieved in the $h$ field. 
This is exactly the case corresponding to the middle panel. We define this critical temperature as {\it the rescue temperature}, $T_{\rm res}$. 
This may lead to the implication whether the transition is successful or not strongly depends on the behavior of  $h$ field evolution. Motivated by this observation, we capture the highest value of the $h$ field at $z=z_0$ starting from $T_c$. The result normalizing to the value of $v_h(T_c)$, $\tilde{h}^{\rm wall}_{\rm max}(T)\equiv h_{\rm max}(z_0,T)/v_h(T_c)$ under different values of $\Delta$ is shown in Fig.~\ref{fig:maxvalue}. 
Treating $\Delta$ as a free parameter, in this analysis we are most interested in how small $\Delta$ is enough to destabilize the domain wall and complete a bubble-free phase transition. 
It is clearly seen from Fig.~\ref{fig:maxvalue} that a good convergence has reached at $\Delta=10^{-2}$, and thus $\Delta =5 \times 10^{-3}$ used in Fig.~\ref{fig:HFevol} is an optimized choice that comprises both efficiency and accuracy of the computation.
Numerically we identify $T_{\rm res}$ as the highest temperature at which $h_{\rm max}(z_0,T)\simeq v_h(T)$ is satisfied, so $T_{\rm res}\simeq71.8{\rm ~GeV}$ for the BMP-A.

In addition to the lattice simulation, $T_{\rm res}$ can be alternatively estimated from the view of energy conservation. 
Based on the above analysis, when the domain wall just becomes unstable, the $s$-field configuration is well described by the field solution, \eq{eq:solution_dw}, and the kinetic energy associated with the domain wall is negligible. 
Let $h_r(z)$ be the $h$ field configuration of the domain wall, then the energy per area deposited into the domain wall, relative to the initial stable wall, is 
\beq
\sigma_V (h_r, T) =\int\! \Big(V\left(0, s_{\rm dw}, T \right) -V\left(h_r, s_{\rm dw},T\right) \Big) d z,
\eeq
here the $z$ integral is performed within the wall region in the vicinity of $z=z_0$. 
At $T\lesssim T_c$, $\sigma_V$ is usually insufficient to overcome the wall tension that is characterized by the gradient energy in the unit area associated with the domain wall, 
\beq
\sigma_g (h_r, T) = \int \frac{1}{2} \left(\partial_z  h_r \right)^2 d z.
\eeq
This results in an oscillatory state in the $h$ field. 
As $T$ decreases, a larger amount of $\sigma_V$ will be deposited into the domain wall. If $\sigma_V$ exceeds $\sigma_g$, this would make possible the destruction of the domain wall. Therefore, we define $T_{\rm res}$ as the highest temperature $T$ that satisfies
\beq
\label{eq:cond}
 g (h_r, T) =0 \,\, \,\, {\rm for}  \,\, \,\, T<T_c,
\eeq
where the function $g$ is given by 
\beq
g (h_r, T) =\sigma_V (h_r, T) - \sigma_g (h_r, T).
\eeq

To find $T_{\rm res}$ using \eq{eq:cond}, the field profile, $h_r(z)$ at $T \simeq T_{\rm res}$ must be known. Observe from Fig.~\ref{fig:HFevol}, at $T_{\rm res}$ the $h$ field configuration appears steady and can be approximately described by a Gaussian wave-packet\footnote{It is equivalent to the lowest Kaluza-Klein state for the $h$ field that was adopted in~\cite{Blasi:2023rqi}.},
\beq
h_{r}(z)=A v_h e^{-\frac{(z-z_0)^2}{\left(\alpha L_{\rm dw}\right)^2}},
\eeq
where $\alpha$ and $A$ are the dimensionless parameters describing the width and the amplitude of the wave-packet, respectively. 
Observing from Fig.~\ref{fig:HFevol}, $A$ is roughly of the order of $\Delta$. Therefore, for the optimal value of $\Delta$ we adopt, the $h^4$ term in \eq{potential} is vanishingly small and can be well dropped. Under this assumption, it is convenient to define a reduced $\tilde{g}$ that is irrelevant to the per-factor $A$, 
\beq
\tilde{g}\left(\alpha, T \right)= g (h_r, T) /A^2.
\eeq

\begin{figure}[t]
\centering
\includegraphics[width=0.45\textwidth]{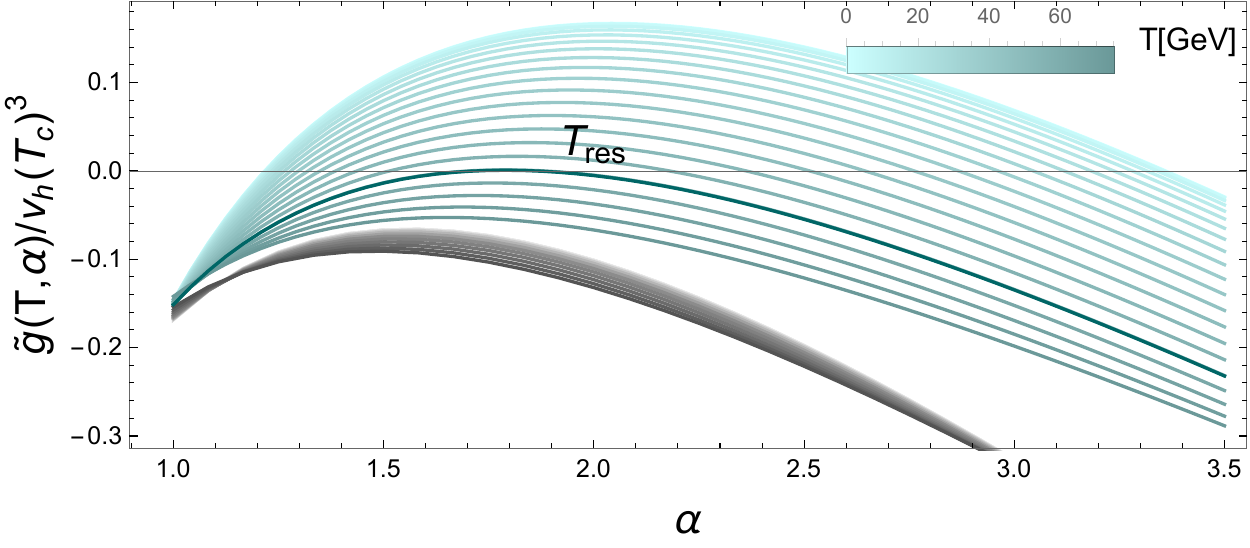} 
\caption{For the BMP-A (cyan) and BMP-B (grey), we generate a bunch of the $\tilde{g}\left(\alpha, T \right)$ curves with the descending value of $T$ from $T_c$ to zero (from bottom to top). BMP-A has a solution for $T_{\rm res}$ indicated in dark line.} 
\label{fig:gfun} 
\end{figure}

We present in Fig.~\ref{fig:gfun} the result of $\tilde{g}$ function for the two BMPs. For each point, we generate a bunch of the $\tilde{g}\left(\alpha, T \right)$ curves at $T$ starting from $T_c$. It turns out that the $\tilde{g}\left(\alpha, T \right)$ function is not invariant with respect to $\alpha$ but, as expected, its amplitude gradually increases over the entire range of $\alpha$ as $T$ decreases. This leads to the practical hint that $T_{\rm res}$, if existing, corresponds to the temperature at which the function $\tilde{g}$ has exactly one zero root. 
Therefore, BMP-A has a solution for $T_{\rm res}$ because there exist $\tilde{g}\left(\alpha, T \right)$ curves having zero roots. 
In contrast, seeding a FOPT without bubbles is impossible in the BMP-B since its $\tilde{g}\left(\alpha, T \right)$ consistently remains negative till the universe cools down to $T=0$.

\begin{table}[t]
\centering
\caption{$T_{\rm res}$ for the BMP-A obtained from various approaches (IF, HF, and TC). The simulation parameters include the size $L$, the lattice number $N^3$, the spacing $\Delta x= L/N$ and $\tilde{n}_{\rm cut}=5$ is taken in the IF. 
In case that the simulation time is insufficient under the requirement $t_{\rm end}/R=1$, a small $L$, corresponding to a small $R$, may predict a lower $T_{\rm res}$ than the actual value (see HF* results). 
}
\begin{tabular}{ l | c c c c c c c c} 
 \hline
 \hline
Methods  & $v_h(T_c)L$ & $N$ & $T_{\rm res} [\text{GeV}]$ &$\sqrt{\langle \delta h^2 \rangle}/v_h(T_c) $
\\ [0.5ex] 
 \hline 
\quad IF &200 & 800    & 71.4& $4\times 10^{-3}$  \cr
\quad HF &800 & 8000       & 71.8& $5\times10^{-3}$\cr 
\quad HF* &200 & 2000       & 69.3& $5\times10^{-3}$\cr 
\quad TC &- & -  & 68.6& - \cr
 \hline
 \hline
\end{tabular}
\label{tab:res}
\end{table}

Finally we compare the results of $T_{\rm res}$ 
obtained using the inhomogeneous fluctuations (IF), the homogeneous fluctuations (HF), and the theoretical calculation (TC) approaches that we have discussed. They are summarized in Table \ref{tab:res}. 
Clearly, the results from the HF and IF methods, under the same magnitude of the fluctuation, have reached satisfactory consistency, although the contribution from $\delta \dot{h}$ is neglected in the HF approach, whereas the TC approach predicts a relatively smaller value. 
As far as we understand, the discrepancy may arise from two factors. In the TC approach, we do not model the $s$-field configuration, which could become a substantial effect for extremely weak couplings. Moreover, we are unable to accurately determine the size of $\alpha$ and, instead, treat it as a free parameter of $\tilde{g}\left(\alpha, T \right)$.

\begin{figure}[t]
\centering 
\includegraphics[width=0.42\textwidth]{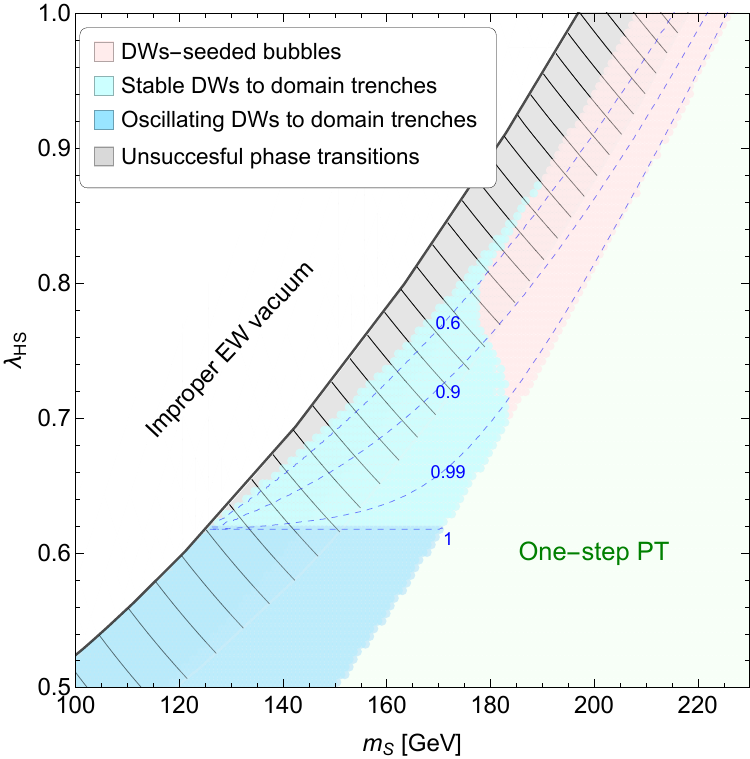} 
\caption{Two-step phase transitions in the model. Homogeneous nucleation is inefficient in the region marked with backslashes. Cyan and blue regions: transition proceeds with the production of domain trenches from domain walls.
Red region: transition proceeds with inhomogeneous bubbles seeded by domain walls. Gray region: eliminated due to unsuccessful phase transition. The blue dashed contours show the value of $T_{\rm res}/T_c$.} 
\label{fig:finalres} 
\end{figure}

\emph{Results \& Discussions}. 
The impact of the mechanism developed in this work is illustrated in Fig.~\ref{fig:finalres} where in the region marked with backslashes a two-step transition is eliminated due to the failure to homogeneous nucleation~\cite{Biekotter:2022kgf}. 
The significant change occurs in the cyan and blue regions where the phase transitions proceed only with the production of domain trenches from domain walls, opening up the new viable parameter region. 
The difference is that in the blue region domain walls are not stable but begin oscillating at $T\gtrsim T_c$, and almost instantaneously become domain trenches at $T_c$, so we estimate $T_{\rm res}\simeq T_c$. 
Outside the backslashed region homogeneous bubbles are efficient, this mechanism still has the potential to cause the transition to occur before the onset of nucleation. 
In the red region this mechanism is also possible but domain trenches occur later than inhomogeneous bubbles, which can seed the phase transition, whereas in the gray region neither mechanism works to complete the phase transition. 
Therefore, the presence of domain walls, although no incontrovertible evidence found in our universe to date, can greatly enrich the way phase transitions are accomplished. 

In general, bubble-free phase transitions can also be achieved in other models with $\mathbb{Z}_2$ domain walls and also other defects. However, two factors should be seriously considered when applying this mechanism. First, if the phase transition is triggered by strong couplings, which typically occurs in models with loop-driven phase transitions, then the initial wall segments around the old domain ($s_\pm$ domains) may spontaneously collapse due to large curvature tension. This will generate the large-sized walls through eating the old domains, or possibly transition directly to the new vacuum ($h$ vacuum) if already developed. 
Second, if significant fluctuations $\delta h$ arises, the $s$-field fluctuation $\delta s$ will be no longer negligible, resulting in an increase in $T_{\rm res}$ or a larger parameter space where the phase transition completes without bubbles. 
Certainly, special attention is needed when the vacuum fluctuations generated during inflation and amplified after horizon exit makes a significant contribution to the $\delta h$ introduced in this work, implying a potential link between inflation physics and new physics at the electroweak scale.

In this mechanism, collisions between domain trenches can generate GWs, and we expect that the power spectrum of the produced GWs is different from those from the traditional FOPT via bubble nucleation. 
If this is true, this would allow us to determine how the phase transition is accomplished through future GW detection.

\section{acknowledgements}
\vspace*{-2mm}
We thank Jose M. No for his useful comments. 
This work is supported by the National Key Research and Development Program of China (Grant No. 2021YFC2203002) and in part by the GuangDong Major Project of Basic and Applied Basic Research (Grant No. 2019B030302001). Y. J. is also funded by the Guangzhou Basic and Applied Basic Research Foundation (No. 202102021092), the GuangDong Basic and Applied Basic Research Foundation (No. 2020A1515110150) and the Sun Yat-sen University Science Foundation.

\bibliography{NoBubblePT} 

\begin{thebibliography}{31}%
\makeatletter
\providecommand \@ifxundefined [1]{%
 \@ifx{#1\undefined}
}%
\providecommand \@ifnum [1]{%
 \ifnum #1\expandafter \@firstoftwo
 \else \expandafter \@secondoftwo
 \fi
}%
\providecommand \@ifx [1]{%
 \ifx #1\expandafter \@firstoftwo
 \else \expandafter \@secondoftwo
 \fi
}%
\providecommand \natexlab [1]{#1}%
\providecommand \enquote  [1]{``#1''}%
\providecommand \bibnamefont  [1]{#1}%
\providecommand \bibfnamefont [1]{#1}%
\providecommand \citenamefont [1]{#1}%
\providecommand \href@noop [0]{\@secondoftwo}%
\providecommand \href [0]{\begingroup \@sanitize@url \@href}%
\providecommand \@href[1]{\@@startlink{#1}\@@href}%
\providecommand \@@href[1]{\endgroup#1\@@endlink}%
\providecommand \@sanitize@url [0]{\catcode `\\12\catcode `\$12\catcode
  `\&12\catcode `\#12\catcode `\^12\catcode `\_12\catcode `\%12\relax}%
\providecommand \@@startlink[1]{}%
\providecommand \@@endlink[0]{}%
\providecommand \url  [0]{\begingroup\@sanitize@url \@url }%
\providecommand \@url [1]{\endgroup\@href {#1}{\urlprefix }}%
\providecommand \urlprefix  [0]{URL }%
\providecommand \Eprint [0]{\href }%
\providecommand \doibase [0]{http://dx.doi.org/}%
\providecommand \selectlanguage [0]{\@gobble}%
\providecommand \bibinfo  [0]{\@secondoftwo}%
\providecommand \bibfield  [0]{\@secondoftwo}%
\providecommand \translation [1]{[#1]}%
\providecommand \BibitemOpen [0]{}%
\providecommand \bibitemStop [0]{}%
\providecommand \bibitemNoStop [0]{.\EOS\space}%
\providecommand \EOS [0]{\spacefactor3000\relax}%
\providecommand \BibitemShut  [1]{\csname bibitem#1\endcsname}%
\let\auto@bib@innerbib\@empty
\bibitem [{\citenamefont {Morrissey}\ and\ \citenamefont
  {Ramsey-Musolf}(2012)}]{Morrissey:2012db}%
  \BibitemOpen
  \bibfield  {author} {\bibinfo {author} {\bibfnamefont {D.~E.}\ \bibnamefont
  {Morrissey}}\ and\ \bibinfo {author} {\bibfnamefont {M.~J.}\ \bibnamefont
  {Ramsey-Musolf}},\ }\href {\doibase 10.1088/1367-2630/14/12/125003}
  {\bibfield  {journal} {\bibinfo  {journal} {New J. Phys.}\ }\textbf {\bibinfo
  {volume} {14}},\ \bibinfo {pages} {125003} (\bibinfo {year} {2012})},\
  \Eprint {http://arxiv.org/abs/1206.2942} {arXiv:1206.2942 [hep-ph]}
  \BibitemShut {NoStop}%
\bibitem [{\citenamefont {Caprini}\ \emph {et~al.}(2020)\citenamefont {Caprini}
  \emph {et~al.}}]{Caprini:2019egz}%
  \BibitemOpen
  \bibfield  {author} {\bibinfo {author} {\bibfnamefont {C.}~\bibnamefont
  {Caprini}} \emph {et~al.},\ }\href {\doibase 10.1088/1475-7516/2020/03/024}
  {\bibfield  {journal} {\bibinfo  {journal} {JCAP}\ }\textbf {\bibinfo
  {volume} {03}},\ \bibinfo {pages} {024} (\bibinfo {year} {2020})},\ \Eprint
  {http://arxiv.org/abs/1910.13125} {arXiv:1910.13125 [astro-ph.CO]}
  \BibitemShut {NoStop}%
\bibitem [{\citenamefont {Mei}\ \emph {et~al.}(2021)\citenamefont {Mei} \emph
  {et~al.}}]{TianQin:2020hid}%
  \BibitemOpen
  \bibfield  {author} {\bibinfo {author} {\bibfnamefont {J.}~\bibnamefont
  {Mei}} \emph {et~al.} (\bibinfo {collaboration} {TianQin}),\ }\href {\doibase
  10.1093/ptep/ptaa114} {\bibfield  {journal} {\bibinfo  {journal} {PTEP}\
  }\textbf {\bibinfo {volume} {2021}},\ \bibinfo {pages} {05A107} (\bibinfo
  {year} {2021})},\ \Eprint {http://arxiv.org/abs/2008.10332} {arXiv:2008.10332
  [gr-qc]} \BibitemShut {NoStop}%
\bibitem [{\citenamefont {Liang}\ \emph {et~al.}(2022)\citenamefont {Liang},
  \citenamefont {Hu}, \citenamefont {Jiang}, \citenamefont {Cheng},
  \citenamefont {Zhang},\ and\ \citenamefont {Mei}}]{Liang:2021bde}%
  \BibitemOpen
  \bibfield  {author} {\bibinfo {author} {\bibfnamefont {Z.-C.}\ \bibnamefont
  {Liang}}, \bibinfo {author} {\bibfnamefont {Y.-M.}\ \bibnamefont {Hu}},
  \bibinfo {author} {\bibfnamefont {Y.}~\bibnamefont {Jiang}}, \bibinfo
  {author} {\bibfnamefont {J.}~\bibnamefont {Cheng}}, \bibinfo {author}
  {\bibfnamefont {J.-d.}\ \bibnamefont {Zhang}}, \ and\ \bibinfo {author}
  {\bibfnamefont {J.}~\bibnamefont {Mei}},\ }\href {\doibase
  10.1103/PhysRevD.105.022001} {\bibfield  {journal} {\bibinfo  {journal}
  {Phys. Rev. D}\ }\textbf {\bibinfo {volume} {105}},\ \bibinfo {pages}
  {022001} (\bibinfo {year} {2022})},\ \Eprint
  {http://arxiv.org/abs/2107.08643} {arXiv:2107.08643 [astro-ph.CO]}
  \BibitemShut {NoStop}%
\bibitem [{\citenamefont {Cai}\ \emph {et~al.}(2017)\citenamefont {Cai},
  \citenamefont {Cao}, \citenamefont {Guo}, \citenamefont {Wang},\ and\
  \citenamefont {Yang}}]{Cai:2017cbj}%
  \BibitemOpen
  \bibfield  {author} {\bibinfo {author} {\bibfnamefont {R.-G.}\ \bibnamefont
  {Cai}}, \bibinfo {author} {\bibfnamefont {Z.}~\bibnamefont {Cao}}, \bibinfo
  {author} {\bibfnamefont {Z.-K.}\ \bibnamefont {Guo}}, \bibinfo {author}
  {\bibfnamefont {S.-J.}\ \bibnamefont {Wang}}, \ and\ \bibinfo {author}
  {\bibfnamefont {T.}~\bibnamefont {Yang}},\ }\href {\doibase
  10.1093/nsr/nwx029} {\bibfield  {journal} {\bibinfo  {journal} {Natl. Sci.
  Rev.}\ }\textbf {\bibinfo {volume} {4}},\ \bibinfo {pages} {687} (\bibinfo
  {year} {2017})},\ \Eprint {http://arxiv.org/abs/1703.00187} {arXiv:1703.00187
  [gr-qc]} \BibitemShut {NoStop}%
\bibitem [{\citenamefont {Baker}\ and\ \citenamefont
  {Kopp}(2017)}]{Baker:2016xzo}%
  \BibitemOpen
  \bibfield  {author} {\bibinfo {author} {\bibfnamefont {M.~J.}\ \bibnamefont
  {Baker}}\ and\ \bibinfo {author} {\bibfnamefont {J.}~\bibnamefont {Kopp}},\
  }\href {\doibase 10.1103/PhysRevLett.119.061801} {\bibfield  {journal}
  {\bibinfo  {journal} {Phys. Rev. Lett.}\ }\textbf {\bibinfo {volume} {119}},\
  \bibinfo {pages} {061801} (\bibinfo {year} {2017})},\ \Eprint
  {http://arxiv.org/abs/1608.07578} {arXiv:1608.07578 [hep-ph]} \BibitemShut
  {NoStop}%
\bibitem [{\citenamefont {Baker}\ \emph {et~al.}(2020)\citenamefont {Baker},
  \citenamefont {Kopp},\ and\ \citenamefont {Long}}]{Baker:2019ndr}%
  \BibitemOpen
  \bibfield  {author} {\bibinfo {author} {\bibfnamefont {M.~J.}\ \bibnamefont
  {Baker}}, \bibinfo {author} {\bibfnamefont {J.}~\bibnamefont {Kopp}}, \ and\
  \bibinfo {author} {\bibfnamefont {A.~J.}\ \bibnamefont {Long}},\ }\href
  {\doibase 10.1103/PhysRevLett.125.151102} {\bibfield  {journal} {\bibinfo
  {journal} {Phys. Rev. Lett.}\ }\textbf {\bibinfo {volume} {125}},\ \bibinfo
  {pages} {151102} (\bibinfo {year} {2020})},\ \Eprint
  {http://arxiv.org/abs/1912.02830} {arXiv:1912.02830 [hep-ph]} \BibitemShut
  {NoStop}%
\bibitem [{\citenamefont {Coleman}(1977)}]{Coleman:1977py}%
  \BibitemOpen
  \bibfield  {author} {\bibinfo {author} {\bibfnamefont {S.~R.}\ \bibnamefont
  {Coleman}},\ }\href {\doibase 10.1103/PhysRevD.16.1248} {\bibfield  {journal}
  {\bibinfo  {journal} {Phys. Rev. D}\ }\textbf {\bibinfo {volume} {15}},\
  \bibinfo {pages} {2929} (\bibinfo {year} {1977})},\ \bibinfo {note}
  {[Erratum: Phys.Rev.D 16, 1248 (1977)]}\BibitemShut {NoStop}%
\bibitem [{\citenamefont {Borrill}\ and\ \citenamefont
  {Gleiser}(1995)}]{Borrill:1994nk}%
  \BibitemOpen
  \bibfield  {author} {\bibinfo {author} {\bibfnamefont {J.}~\bibnamefont
  {Borrill}}\ and\ \bibinfo {author} {\bibfnamefont {M.}~\bibnamefont
  {Gleiser}},\ }\href {\doibase 10.1103/PhysRevD.51.4111} {\bibfield  {journal}
  {\bibinfo  {journal} {Phys. Rev. D}\ }\textbf {\bibinfo {volume} {51}},\
  \bibinfo {pages} {4111} (\bibinfo {year} {1995})},\ \Eprint
  {http://arxiv.org/abs/hep-ph/9410235} {arXiv:hep-ph/9410235} \BibitemShut
  {NoStop}%
\bibitem [{\citenamefont {Anderson}\ and\ \citenamefont
  {Hall}(1992)}]{Anderson:1991zb}%
  \BibitemOpen
  \bibfield  {author} {\bibinfo {author} {\bibfnamefont {G.~W.}\ \bibnamefont
  {Anderson}}\ and\ \bibinfo {author} {\bibfnamefont {L.~J.}\ \bibnamefont
  {Hall}},\ }\href {\doibase 10.1103/PhysRevD.45.2685} {\bibfield  {journal}
  {\bibinfo  {journal} {Phys. Rev. D}\ }\textbf {\bibinfo {volume} {45}},\
  \bibinfo {pages} {2685} (\bibinfo {year} {1992})}\BibitemShut {NoStop}%
\bibitem [{\citenamefont {Guth}\ and\ \citenamefont
  {Weinberg}(1983)}]{Guth:1982pn}%
  \BibitemOpen
  \bibfield  {author} {\bibinfo {author} {\bibfnamefont {A.~H.}\ \bibnamefont
  {Guth}}\ and\ \bibinfo {author} {\bibfnamefont {E.~J.}\ \bibnamefont
  {Weinberg}},\ }\href {\doibase 10.1016/0550-3213(83)90307-3} {\bibfield
  {journal} {\bibinfo  {journal} {Nucl. Phys. B}\ }\textbf {\bibinfo {volume}
  {212}},\ \bibinfo {pages} {321} (\bibinfo {year} {1983})}\BibitemShut
  {NoStop}%
\bibitem [{\citenamefont {Hawking}\ \emph {et~al.}(1982)\citenamefont
  {Hawking}, \citenamefont {Moss},\ and\ \citenamefont
  {Stewart}}]{Hawking:1982ga}%
  \BibitemOpen
  \bibfield  {author} {\bibinfo {author} {\bibfnamefont {S.~W.}\ \bibnamefont
  {Hawking}}, \bibinfo {author} {\bibfnamefont {I.~G.}\ \bibnamefont {Moss}}, \
  and\ \bibinfo {author} {\bibfnamefont {J.~M.}\ \bibnamefont {Stewart}},\
  }\href {\doibase 10.1103/PhysRevD.26.2681} {\bibfield  {journal} {\bibinfo
  {journal} {Phys. Rev. D}\ }\textbf {\bibinfo {volume} {26}},\ \bibinfo
  {pages} {2681} (\bibinfo {year} {1982})}\BibitemShut {NoStop}%
\bibitem [{\citenamefont {Baum}\ \emph {et~al.}(2021)\citenamefont {Baum},
  \citenamefont {Carena}, \citenamefont {Shah}, \citenamefont {Wagner},\ and\
  \citenamefont {Wang}}]{Baum:2020vfl}%
  \BibitemOpen
  \bibfield  {author} {\bibinfo {author} {\bibfnamefont {S.}~\bibnamefont
  {Baum}}, \bibinfo {author} {\bibfnamefont {M.}~\bibnamefont {Carena}},
  \bibinfo {author} {\bibfnamefont {N.~R.}\ \bibnamefont {Shah}}, \bibinfo
  {author} {\bibfnamefont {C.~E.~M.}\ \bibnamefont {Wagner}}, \ and\ \bibinfo
  {author} {\bibfnamefont {Y.}~\bibnamefont {Wang}},\ }\href {\doibase
  10.1007/JHEP03(2021)055} {\bibfield  {journal} {\bibinfo  {journal} {JHEP}\
  }\textbf {\bibinfo {volume} {03}},\ \bibinfo {pages} {055} (\bibinfo {year}
  {2021})},\ \Eprint {http://arxiv.org/abs/2009.10743} {arXiv:2009.10743
  [hep-ph]} \BibitemShut {NoStop}%
\bibitem [{\citenamefont {Biek\"otter}\ \emph {et~al.}(2021)\citenamefont
  {Biek\"otter}, \citenamefont {Heinemeyer}, \citenamefont {No}, \citenamefont
  {Olea},\ and\ \citenamefont {Weiglein}}]{Biekotter:2021ysx}%
  \BibitemOpen
  \bibfield  {author} {\bibinfo {author} {\bibfnamefont {T.}~\bibnamefont
  {Biek\"otter}}, \bibinfo {author} {\bibfnamefont {S.}~\bibnamefont
  {Heinemeyer}}, \bibinfo {author} {\bibfnamefont {J.~M.}\ \bibnamefont {No}},
  \bibinfo {author} {\bibfnamefont {M.~O.}\ \bibnamefont {Olea}}, \ and\
  \bibinfo {author} {\bibfnamefont {G.}~\bibnamefont {Weiglein}},\ }\href
  {\doibase 10.1088/1475-7516/2021/06/018} {\bibfield  {journal} {\bibinfo
  {journal} {JCAP}\ }\textbf {\bibinfo {volume} {06}},\ \bibinfo {pages} {018}
  (\bibinfo {year} {2021})},\ \Eprint {http://arxiv.org/abs/2103.12707}
  {arXiv:2103.12707 [hep-ph]} \BibitemShut {NoStop}%
\bibitem [{\citenamefont {Gon\c{c}alves}\ \emph {et~al.}(2022)\citenamefont
  {Gon\c{c}alves}, \citenamefont {Kaladharan},\ and\ \citenamefont
  {Wu}}]{Goncalves:2021egx}%
  \BibitemOpen
  \bibfield  {author} {\bibinfo {author} {\bibfnamefont {D.}~\bibnamefont
  {Gon\c{c}alves}}, \bibinfo {author} {\bibfnamefont {A.}~\bibnamefont
  {Kaladharan}}, \ and\ \bibinfo {author} {\bibfnamefont {Y.}~\bibnamefont
  {Wu}},\ }\href {\doibase 10.1103/PhysRevD.105.095041} {\bibfield  {journal}
  {\bibinfo  {journal} {Phys. Rev. D}\ }\textbf {\bibinfo {volume} {105}},\
  \bibinfo {pages} {095041} (\bibinfo {year} {2022})},\ \Eprint
  {http://arxiv.org/abs/2108.05356} {arXiv:2108.05356 [hep-ph]} \BibitemShut
  {NoStop}%
\bibitem [{\citenamefont {Biek\"otter}\ \emph {et~al.}(2023)\citenamefont
  {Biek\"otter}, \citenamefont {Heinemeyer}, \citenamefont {No}, \citenamefont
  {Olea-Romacho},\ and\ \citenamefont {Weiglein}}]{Biekotter:2022kgf}%
  \BibitemOpen
  \bibfield  {author} {\bibinfo {author} {\bibfnamefont {T.}~\bibnamefont
  {Biek\"otter}}, \bibinfo {author} {\bibfnamefont {S.}~\bibnamefont
  {Heinemeyer}}, \bibinfo {author} {\bibfnamefont {J.~M.}\ \bibnamefont {No}},
  \bibinfo {author} {\bibfnamefont {M.~O.}\ \bibnamefont {Olea-Romacho}}, \
  and\ \bibinfo {author} {\bibfnamefont {G.}~\bibnamefont {Weiglein}},\ }\href
  {\doibase 10.1088/1475-7516/2023/03/031} {\bibfield  {journal} {\bibinfo
  {journal} {JCAP}\ }\textbf {\bibinfo {volume} {03}},\ \bibinfo {pages} {031}
  (\bibinfo {year} {2023})},\ \Eprint {http://arxiv.org/abs/2208.14466}
  {arXiv:2208.14466 [hep-ph]} \BibitemShut {NoStop}%
\bibitem [{\citenamefont {Vilenkin}\ and\ \citenamefont
  {Shellard}(2000)}]{Vilenkin:2000jqa}%
  \BibitemOpen
  \bibfield  {author} {\bibinfo {author} {\bibfnamefont {A.}~\bibnamefont
  {Vilenkin}}\ and\ \bibinfo {author} {\bibfnamefont {E.~P.~S.}\ \bibnamefont
  {Shellard}},\ }\href@noop {} {\emph {\bibinfo {title} {{Cosmic Strings and
  Other Topological Defects}}}}\ (\bibinfo  {publisher} {Cambridge University
  Press},\ \bibinfo {year} {2000})\BibitemShut {NoStop}%
\bibitem [{\citenamefont {Blasi}\ and\ \citenamefont
  {Mariotti}(2022)}]{Blasi:2022woz}%
  \BibitemOpen
  \bibfield  {author} {\bibinfo {author} {\bibfnamefont {S.}~\bibnamefont
  {Blasi}}\ and\ \bibinfo {author} {\bibfnamefont {A.}~\bibnamefont
  {Mariotti}},\ }\href {\doibase 10.1103/PhysRevLett.129.261303} {\bibfield
  {journal} {\bibinfo  {journal} {Phys. Rev. Lett.}\ }\textbf {\bibinfo
  {volume} {129}},\ \bibinfo {pages} {261303} (\bibinfo {year} {2022})},\
  \Eprint {http://arxiv.org/abs/2203.16450} {arXiv:2203.16450 [hep-ph]}
  \BibitemShut {NoStop}%
\bibitem [{\citenamefont {Blasi}\ \emph {et~al.}(2023)\citenamefont {Blasi},
  \citenamefont {Jinno}, \citenamefont {Konstandin}, \citenamefont {Rubira},\
  and\ \citenamefont {Stomberg}}]{Blasi:2023rqi}%
  \BibitemOpen
  \bibfield  {author} {\bibinfo {author} {\bibfnamefont {S.}~\bibnamefont
  {Blasi}}, \bibinfo {author} {\bibfnamefont {R.}~\bibnamefont {Jinno}},
  \bibinfo {author} {\bibfnamefont {T.}~\bibnamefont {Konstandin}}, \bibinfo
  {author} {\bibfnamefont {H.}~\bibnamefont {Rubira}}, \ and\ \bibinfo {author}
  {\bibfnamefont {I.}~\bibnamefont {Stomberg}},\ }\href {\doibase
  10.1088/1475-7516/2023/10/051} {\bibfield  {journal} {\bibinfo  {journal}
  {JCAP}\ }\textbf {\bibinfo {volume} {10}},\ \bibinfo {pages} {051} (\bibinfo
  {year} {2023})},\ \Eprint {http://arxiv.org/abs/2302.06952} {arXiv:2302.06952
  [astro-ph.CO]} \BibitemShut {NoStop}%
\bibitem [{\citenamefont {Agrawal}\ \emph {et~al.}(2023)\citenamefont
  {Agrawal}, \citenamefont {Blasi}, \citenamefont {Mariotti},\ and\
  \citenamefont {Nee}}]{Agrawal:2023cgp}%
  \BibitemOpen
  \bibfield  {author} {\bibinfo {author} {\bibfnamefont {P.}~\bibnamefont
  {Agrawal}}, \bibinfo {author} {\bibfnamefont {S.}~\bibnamefont {Blasi}},
  \bibinfo {author} {\bibfnamefont {A.}~\bibnamefont {Mariotti}}, \ and\
  \bibinfo {author} {\bibfnamefont {M.}~\bibnamefont {Nee}},\ }\href@noop {} {\
   (\bibinfo {year} {2023})},\ \Eprint {http://arxiv.org/abs/2312.06749}
  {arXiv:2312.06749 [hep-ph]} \BibitemShut {NoStop}%
\bibitem [{\citenamefont {Zeldovich}\ \emph {et~al.}(1974)\citenamefont
  {Zeldovich}, \citenamefont {Kobzarev},\ and\ \citenamefont
  {Okun}}]{Zeldovich:1974uw}%
  \BibitemOpen
  \bibfield  {author} {\bibinfo {author} {\bibfnamefont {Y.~B.}\ \bibnamefont
  {Zeldovich}}, \bibinfo {author} {\bibfnamefont {I.~Y.}\ \bibnamefont
  {Kobzarev}}, \ and\ \bibinfo {author} {\bibfnamefont {L.~B.}\ \bibnamefont
  {Okun}},\ }\href@noop {} {\bibfield  {journal} {\bibinfo  {journal} {Zh.
  Eksp. Teor. Fiz.}\ }\textbf {\bibinfo {volume} {67}},\ \bibinfo {pages} {3}
  (\bibinfo {year} {1974})}\BibitemShut {NoStop}%
\bibitem [{\citenamefont {Tumasyan}\ \emph {et~al.}(2022)\citenamefont
  {Tumasyan} \emph {et~al.}}]{CMS:2022dwd}%
  \BibitemOpen
  \bibfield  {author} {\bibinfo {author} {\bibfnamefont {A.}~\bibnamefont
  {Tumasyan}} \emph {et~al.} (\bibinfo {collaboration} {CMS}),\ }\href
  {\doibase 10.1038/s41586-022-04892-x} {\bibfield  {journal} {\bibinfo
  {journal} {Nature}\ }\textbf {\bibinfo {volume} {607}},\ \bibinfo {pages}
  {60} (\bibinfo {year} {2022})},\ \Eprint {http://arxiv.org/abs/2207.00043}
  {arXiv:2207.00043 [hep-ex]} \BibitemShut {NoStop}%
\bibitem [{\citenamefont {Aad}\ \emph {et~al.}(2022)\citenamefont {Aad} \emph
  {et~al.}}]{ATLAS:2022vkf}%
  \BibitemOpen
  \bibfield  {author} {\bibinfo {author} {\bibfnamefont {G.}~\bibnamefont
  {Aad}} \emph {et~al.} (\bibinfo {collaboration} {ATLAS}),\ }\href {\doibase
  10.1038/s41586-022-04893-w} {\bibfield  {journal} {\bibinfo  {journal}
  {Nature}\ }\textbf {\bibinfo {volume} {607}},\ \bibinfo {pages} {52}
  (\bibinfo {year} {2022})},\ \bibinfo {note} {[Erratum: Nature 612, E24
  (2022)]},\ \Eprint {http://arxiv.org/abs/2207.00092} {arXiv:2207.00092
  [hep-ex]} \BibitemShut {NoStop}%
\bibitem [{\citenamefont {Espinosa}\ \emph {et~al.}(2012)\citenamefont
  {Espinosa}, \citenamefont {Konstandin},\ and\ \citenamefont
  {Riva}}]{Espinosa:2011ax}%
  \BibitemOpen
  \bibfield  {author} {\bibinfo {author} {\bibfnamefont {J.~R.}\ \bibnamefont
  {Espinosa}}, \bibinfo {author} {\bibfnamefont {T.}~\bibnamefont
  {Konstandin}}, \ and\ \bibinfo {author} {\bibfnamefont {F.}~\bibnamefont
  {Riva}},\ }\href {\doibase 10.1016/j.nuclphysb.2011.09.010} {\bibfield
  {journal} {\bibinfo  {journal} {Nucl. Phys. B}\ }\textbf {\bibinfo {volume}
  {854}},\ \bibinfo {pages} {592} (\bibinfo {year} {2012})},\ \Eprint
  {http://arxiv.org/abs/1107.5441} {arXiv:1107.5441 [hep-ph]} \BibitemShut
  {NoStop}%
\bibitem [{\citenamefont {Saikawa}(2017)}]{Saikawa:2017hiv}%
  \BibitemOpen
  \bibfield  {author} {\bibinfo {author} {\bibfnamefont {K.}~\bibnamefont
  {Saikawa}},\ }\href {\doibase 10.3390/universe3020040} {\bibfield  {journal}
  {\bibinfo  {journal} {Universe}\ }\textbf {\bibinfo {volume} {3}},\ \bibinfo
  {pages} {40} (\bibinfo {year} {2017})},\ \Eprint
  {http://arxiv.org/abs/1703.02576} {arXiv:1703.02576 [hep-ph]} \BibitemShut
  {NoStop}%
\bibitem [{\citenamefont {Braden}\ \emph {et~al.}(2019)\citenamefont {Braden},
  \citenamefont {Johnson}, \citenamefont {Peiris}, \citenamefont {Pontzen},\
  and\ \citenamefont {Weinfurtner}}]{Braden:2018tky}%
  \BibitemOpen
  \bibfield  {author} {\bibinfo {author} {\bibfnamefont {J.}~\bibnamefont
  {Braden}}, \bibinfo {author} {\bibfnamefont {M.~C.}\ \bibnamefont {Johnson}},
  \bibinfo {author} {\bibfnamefont {H.~V.}\ \bibnamefont {Peiris}}, \bibinfo
  {author} {\bibfnamefont {A.}~\bibnamefont {Pontzen}}, \ and\ \bibinfo
  {author} {\bibfnamefont {S.}~\bibnamefont {Weinfurtner}},\ }\href {\doibase
  10.1103/PhysRevLett.123.031601} {\bibfield  {journal} {\bibinfo  {journal}
  {Phys. Rev. Lett.}\ }\textbf {\bibinfo {volume} {123}},\ \bibinfo {pages}
  {031601} (\bibinfo {year} {2019})},\ \bibinfo {note} {[Erratum:
  Phys.Rev.Lett. 129, 059901 (2022)]},\ \Eprint
  {http://arxiv.org/abs/1806.06069} {arXiv:1806.06069 [hep-th]} \BibitemShut
  {NoStop}%
\bibitem [{\citenamefont {Liddle}\ and\ \citenamefont
  {Lyth}(2000)}]{Liddle:2000cg}%
  \BibitemOpen
  \bibfield  {author} {\bibinfo {author} {\bibfnamefont {A.~R.}\ \bibnamefont
  {Liddle}}\ and\ \bibinfo {author} {\bibfnamefont {D.~H.}\ \bibnamefont
  {Lyth}},\ }\href {\doibase 10.1017/CBO9781139175180} {\emph {\bibinfo {title}
  {{Cosmological inflation and large scale structure}}}}\ (\bibinfo {year}
  {2000})\BibitemShut {NoStop}%
\bibitem [{\citenamefont {Press}\ \emph {et~al.}(1989)\citenamefont {Press},
  \citenamefont {Ryden},\ and\ \citenamefont {Spergel}}]{Press:1989yh}%
  \BibitemOpen
  \bibfield  {author} {\bibinfo {author} {\bibfnamefont {W.~H.}\ \bibnamefont
  {Press}}, \bibinfo {author} {\bibfnamefont {B.~S.}\ \bibnamefont {Ryden}}, \
  and\ \bibinfo {author} {\bibfnamefont {D.~N.}\ \bibnamefont {Spergel}},\
  }\href {\doibase 10.1086/168151} {\bibfield  {journal} {\bibinfo  {journal}
  {Astrophys. J.}\ }\textbf {\bibinfo {volume} {347}},\ \bibinfo {pages} {590}
  (\bibinfo {year} {1989})}\BibitemShut {NoStop}%
\bibitem [{\citenamefont {Felder}\ and\ \citenamefont
  {Tkachev}(2008)}]{Felder:2000hq}%
  \BibitemOpen
  \bibfield  {author} {\bibinfo {author} {\bibfnamefont {G.~N.}\ \bibnamefont
  {Felder}}\ and\ \bibinfo {author} {\bibfnamefont {I.}~\bibnamefont
  {Tkachev}},\ }\href {\doibase 10.1016/j.cpc.2008.02.009} {\bibfield
  {journal} {\bibinfo  {journal} {Comput. Phys. Commun.}\ }\textbf {\bibinfo
  {volume} {178}},\ \bibinfo {pages} {929} (\bibinfo {year} {2008})},\ \Eprint
  {http://arxiv.org/abs/hep-ph/0011159} {arXiv:hep-ph/0011159} \BibitemShut
  {NoStop}%
\bibitem [{\citenamefont {Figueroa}\ \emph {et~al.}(2021)\citenamefont
  {Figueroa}, \citenamefont {Florio}, \citenamefont {Torrenti},\ and\
  \citenamefont {Valkenburg}}]{Figueroa:2020rrl}%
  \BibitemOpen
  \bibfield  {author} {\bibinfo {author} {\bibfnamefont {D.~G.}\ \bibnamefont
  {Figueroa}}, \bibinfo {author} {\bibfnamefont {A.}~\bibnamefont {Florio}},
  \bibinfo {author} {\bibfnamefont {F.}~\bibnamefont {Torrenti}}, \ and\
  \bibinfo {author} {\bibfnamefont {W.}~\bibnamefont {Valkenburg}},\ }\href
  {\doibase 10.1088/1475-7516/2021/04/035} {\bibfield  {journal} {\bibinfo
  {journal} {JCAP}\ }\textbf {\bibinfo {volume} {04}},\ \bibinfo {pages} {035}
  (\bibinfo {year} {2021})},\ \Eprint {http://arxiv.org/abs/2006.15122}
  {arXiv:2006.15122 [astro-ph.CO]} \BibitemShut {NoStop}%
\bibitem [{\citenamefont {Press}\ \emph {et~al.}(1992)\citenamefont {Press},
  \citenamefont {Teukolsky}, \citenamefont {Vetterling},\ and\ \citenamefont
  {Flannery}}]{Press:1992zz}%
  \BibitemOpen
  \bibfield  {author} {\bibinfo {author} {\bibfnamefont {W.~H.}\ \bibnamefont
  {Press}}, \bibinfo {author} {\bibfnamefont {S.~A.}\ \bibnamefont
  {Teukolsky}}, \bibinfo {author} {\bibfnamefont {W.~T.}\ \bibnamefont
  {Vetterling}}, \ and\ \bibinfo {author} {\bibfnamefont {B.~P.}\ \bibnamefont
  {Flannery}},\ }\href {\doibase 10.1111/j.1539-6924.1989.tb01007.x} {\emph
  {\bibinfo {title} {{Numerical Recipes in C (2nd ed.): The Art of Scientific
  Computing}}}}\ (\bibinfo {year} {1992})\BibitemShut {NoStop}%
\end{thebibliography}%

\appendix

\section{A. Linearized equations for the perturbed fields}

In general, any scalar field $\phi(\mathbf{x}, t)$ (for instance, $\phi=h,s$) is composed of the homogeneous background field $\phi_0(t)$ and perturbation $\delta \phi (\mathbf{x}, t)$,
\beq
\label{eq:phiexpand}
\phi(\mathbf{x}, t)=\phi_0(t)+\delta \phi(\mathbf{x},t), 
\eeq
Here $\phi_0(t)$ is the homogeneous classical solution to the field equations and $\delta \phi (\mathbf{x}, t)$ amounts to quantum fluctuation on the field. We work within the linearized theory, assuming that the perturbation $\delta \phi (\mathbf{x}, t)$ is small. 

For the potential $V(h,s)$ involving two scalar fields, we expand it as, to the first order, 
\beq
\label{eq:Vexpand}
V(h, s)= V\left(h_0, s_0\right)+ V_h^{\prime}\left(h_0, s_0\right) \delta h + V_s^{\prime}\left(h_0, s_0\right) \delta s +\dots \\
\eeq
where $V_\phi^{\prime} (h_0,s_0)\equiv {\partial V(h,s) \over \partial \phi} \big|_{h=h_0,s=s_0}$ and $\dots$ denotes the terms containing more than one perturbed field give the various source of the back-reactions, which are not of our interest in the analysis. 

Taking the derivative of \eq{eq:Vexpand} with respect to each field yields,
\beq
\label{eq:dVexpand}
\begin{aligned}
\partial_h V(h, s)= V_{hh}^{\prime\prime}\left(h_0, s_0\right) \delta h + V_{hs}^{\prime\prime}\left(h_0, s_0\right) \delta s +\dots \\
\partial_s V(h, s)= V_{ss}^{\prime\prime}\left(h_0, s_0\right) \delta s + V_{hs}^{\prime\prime}\left(h_0, s_0\right) \delta h + \dots 
\end{aligned}
\eeq

Taking advantage of $V_{\{h,s\}}^{\prime}\left(0, s_0\right)=0$ attributed to the symmetric potential. 
Substituting \eq{eq:Vexpand} and \eq{eq:phiexpand} into the Klein-Gordon equation in the flat metric
\beq
\label{eq:KG}
\ddot{\phi}-\nabla^2 \phi+\frac{\partial V(h,s)}{\partial  \phi}=0, 
\eeq
yields the equations for the background fields $\phi_0=h_0,s_0$
\beq
\label{eq:bkgsol}
\ddot{\phi_0}(t)+V_\phi^{\prime}(h_0,s_0)=0,
\eeq
and the one for the perturbed fields
\beq
\label{eq:pertsol}
\left(\partial^2_t -\nabla^2 + V^{\prime\prime}_{\phi\phi} (h_0,s_0)\right){\delta\phi}(\mathbf{x}, t) =0.
\eeq

Writing the Fourier series for the perturbed fields
\beq
{\delta\phi}(\mathbf{x}, t)=\int {d^3 k\over(2\pi)^3} \delta {\tilde \phi}_{\bf k} e^{i{\bf k}\cdot {\bf x}}, 
\eeq
the Fourier-transformed mode $\delta {\tilde \phi}_{\bf k}$, for a given ${\bf k}$, obeys the linearized equation 
\beq
(\partial^2_t + \omega^2_{\bf k} )\delta {\tilde \phi}_{\bf k}(t)=0,
\label{eq:lineom}
\eeq
where $\omega^{2}_{\bf k}=|{\bf k}|^2+V^{\prime \prime}_{\phi\phi}(h_0, s_0)$ evaluated at the vacuum $(h_0, s_0)$ for the last term.

\section{B. The initialization of the field configuration} 
\label{app:lattice}
In the simulation, we consider a box with the size of $L$ and $N=800$ points in each dimension, so the x-lattice spacing is $\epsilon = L/N$. The former parameter determines the spacing of the conjugated k-lattice, $\tilde{\epsilon}=2\pi /L$ while the latter constrains the size of k-lattice with the allowed maximum $k$-value, $k_{\rm max}=\tilde{\epsilon}N/2=\pi N/L$. For later convenience, we label the site of the x-lattice with ${\bf n}\epsilon=\{n_x,n_y, n_z\} \epsilon$, with $n_i=0,1, \dots, N-1$ and of the k-lattice with $\tilde{\bf n}\tilde{\epsilon}=\{\tilde{n}_x,\tilde{n}_y,\tilde{n}_z\} \tilde{\epsilon}$ with $\tilde{n}_i=(-N/2+1),(-N/2+2),\dots,0\dots,N/2$.   
For the sake of energy conservation, the period conditions for the field configuration on the boundaries are imposed. 
To fulfill this requirement, we model a pair of domain walls that are localized at $\pm z_0$ using 
 \beq
 s({\bf x} , T) = s_{\rm dw}({\bf x}, z_0, T)-s_{\rm dw}({\bf x}, -z_0, T)+v_s(T),
 \eeq
with $s_{\rm dw}({\bf x}, z_0, T)$ being the profile for each domain wall given in \eq{eq:solution_dw}, 
so the $s$-field configuration is initialized on the x-lattice. 
 
However, it is difficult to directly establish the $h$-field configuration on the x-lattice. 
Taking advantage of the relation $\delta \tilde{h}_{-{\bf k}}=\delta \tilde{h}^*_{\bf k}$ it is sufficient to consider $\tilde{n_i}\ge0$ sites. 
For the Gaussian field $\delta\tilde{h}_{\bf k}$, \eq{eq:distr}, one can generate the initial configuration using the Box-Muller method~\cite{Felder:2000hq, Figueroa:2020rrl},
\beq
\label{hk}
\delta\tilde{h}_{\bf k} =L^{3 \over 2}e^{i 2\pi \alpha}\sqrt{-\sigma^2_k\ln \beta},
\eeq
where $\alpha, \beta$ are two irrelevant random numbers within $(0,1)$ and the pre-factor $L^{3 \over 2}$ is introduced from naive dimensional analysis. 
Apparently, $\delta\tilde{h}_{\bf k}$ has a spherical symmetry and its discretized form on the k-lattice is 
 \beq
 \delta \tilde{h}_{\tilde{\bf n}}=L^{\frac{3}{2}}e^{i 2\pi \alpha }\sqrt{-\sigma^2_{\tilde{n}}\ln \beta },
 \label{hk}
 \eeq
 where the random numbers $\alpha,\beta$ are independently selected on each site of the k-lattice labeled by $\tilde{\bf n}\tilde{\epsilon}=\{\tilde{n}_x,\tilde{n}_y,\tilde{n}_z\} \tilde{\epsilon}$ and $\sigma^2_{\tilde{n}}=(2\sqrt{(\tilde{n}\tilde{\epsilon})^2+V^{\prime \prime}_{hh}(0,\pm v_s)})^{-1}$ with $\tilde{n}\equiv |\tilde{\bf n}|$. 
 In our tests the stochastic error in the variance of $ \delta \tilde{h}_{\tilde{\bf n}}$ distribution is as small as milli-percent level. 
   
Once $\delta \tilde{h}_{\tilde{\bf n}}$ is initialized on the k-lattice, we can generate the field configuration $\delta h$ on the x-lattice via the discretized inverse Fourier transformation,
\beq
\label{eq:deltah}
\delta h({\bf n})= \frac{1}{L^3}  \sum_{\tilde{\bf n}} e^{-i\frac{2 \pi }{N} {\tilde{\bf n}} \cdot {\bf n}}  \delta \tilde{h}_{\tilde{\bf n}},
\eeq
which gives the ensemble average of $\delta h^2({\bf n})$,
\beq
\label{eq:rms}
\langle \delta h^2({\bf n}) \rangle_{V}=\frac{\tilde{\epsilon}^3}{(2\pi)^3 L^3} \sum_{\tilde{\bf n}}  | \delta {\tilde h}_{\tilde{\bf n}}|^2.
\eeq

Because of the finite size sampled in the k-lattice, the summation over the sites in a spherical volume is required in generating \eq{eq:deltah} to avoid signal distortion. 
According to the Nyquist's theorem~\cite{Press:1992zz}, the size of the volume must not exceed the highest value of $\tilde {n}$ available in the entire $k$-lattice, which means $\tilde{n}_{\rm max}=k_{\rm max}/\tilde{\epsilon}=400$ in our case. 
Therefore, in practice it is sufficient to perform the summation over the sites such that $ |\tilde{\bf n}| \leq \tilde{n}_{\rm max}$.
 
\begin{table}[b]
\centering
\caption{The root mean square of $\delta h({\bf n})$ normalized to $v_h(T_c)$  for the BMP-A. The simulation parameters $v_h(T_c)L=200$ and  $N=800$ are taken.}
\begin{tabular}{ c | c c c c c c c c} 
 \hline
 \hline
$\tilde{n}_{\rm cut}$  & 5 & 15 & 25 & 50 & 100 \cr
 \hline 
 $\sqrt{\langle \delta h^2 \rangle}/v_h(T_c) $ & 0.004 & 0.02 & 0.04 & 0.09 & 0.19 \cr
 \hline
 \hline
\end{tabular}
\label{tab:ncut}
\end{table}

On the other hand, from \eq{eq:rms} we see that the mean square diverges because $| \delta {\tilde h}_{\tilde{\bf n}}|^2$ in the sum fails to vanish in the limit of large $\tilde{\bf n}$ (also large $k$). In cosmology such divergence is actually an indication of rich structure on small scales, which can be get rid of by smoothing~\cite{Liddle:2000cg} or simply introducing an ultra-violet cutoff $\tilde{n}_{\rm cut}$. The choice of $\tilde{n}_{\rm cut}$ involves the numerical convergence. In principle, it is free to choose any value below $\tilde{n}_{\rm max}$, but this will directly determine the magnitude of the mean square. 
The normalized root mean square of $\delta h({\bf n})$ generated by different $\tilde{n}_{\rm cut}$ is given in Table~\ref{tab:ncut}. 
Since the purpose of our work is to study the extent to which the vacuum fluctuations destabilize domain walls and complete bubble-free phase transitions, we avoid introducing too large fluctuations and thus $\tilde{n}_{\rm cut}=5$, as an example of small fluctuation scenarios, is adopted in the IF approach.

\end{document}